\documentclass[a4paper]{article}
\usepackage{amsmath,graphicx,amssymb,delarray,subfigure,verbatim, bbding}
\usepackage{diagbox, makecell, mathtools}
\usepackage{bbding}
\usepackage{diagbox, makecell}
\usepackage{multirow}
\usepackage{mathrsfs}
\usepackage{booktabs}
\usepackage{INTERSPEECH2022}
\usepackage{threeparttable}

\makeatletter
\renewcommand\normalsize{%
	\abovedisplayskip 3\p@ \@plus3\p@ \@minus3\p@
	\abovedisplayshortskip \z@ \@plus3\p@
	\belowdisplayshortskip 6\p@ \@plus3\p@ \@minus3\p@ 
	\belowdisplayskip \abovedisplayskip
	\let\@listi\@listI}
\makeatother

\title{TMGAN-PLC: Audio Packet Loss Concealment using Temporal Memory Generative Adversarial Network}
\name{Yuansheng Guan$^{1, 2}$, Guochen Yu$^{2, 3}$, Andong Li$^{2}$, Chengshi Zheng$^{2,*}$ \thanks{Chengshi Zheng is the corresponding author.}, Jie Wang$^{1}$}
\address{
	$^1$School of Electronics and Communication Engineering, Guangzhou University, Guangzhou, China\\
	$^2$Key Laboratory of Noise and Vibration Research, Institute of Acoustics, Chinese Academy of Sciences, Beijing, China\\
	$^3$Communication University of China, Beijing, China
}
\email{2111907059@e.gzhu.edu.cn, \{yuguochen\}@cuc.edu.cn, \{liandong, cszheng\}@mail.ioa.ac.cn, wangjie@gzhu.edu.cn}
\begin{document}
\maketitle
\begin{abstract}
\vspace{-0.2cm}
Real-time communications in packet-switched networks have become widely used in daily communication, while they inevitably suffer from network delays and data losses in constrained real-time conditions. To solve these problems, audio packet loss concealment (PLC) algorithms have been developed to mitigate voice transmission failures by reconstructing the lost information. Limited by the transmission latency and device memory, it is still intractable for PLC to accomplish high-quality voice reconstruction using a relatively small packet buffer. In this paper, we propose a temporal memory generative adversarial network for audio PLC, dubbed TMGAN-PLC, which is comprised of a novel nested-UNet generator and the time-domain/frequency-domain discriminators. Specifically, a combination of the nested-UNet and temporal feature-wise linear modulation is elaborately devised in the generator to finely adjust the intra-frame information and establish inter-frame temporal dependencies. To complement the missing speech content caused by longer loss bursts, we employ multi-stage gated vector quantizers to capture the correct content and reconstruct the near-real smooth audio. Extensive experiments on the PLC Challenge dataset demonstrate that the proposed method yields promising performance in terms of speech quality, intelligibility, and PLCMOS. 
	
\end{abstract}
\vspace{-0.3cm}
\section{Introduction}
\vspace{-0.2cm}
With the end-to-end digital packet switched telephone system becoming increasingly important in daily communication, the degradation of transmitted audio quality has become an endogenous problem to be solved, which is typically caused by packet loss, delay, and unrecoverable bit errors during transmission~{\cite{takahashi2004packet}}. The receiver outputs silence when packets are not received properly, and audible distortion can be identified even at very low packet-loss rates. In this respect, "Packet Loss Concealment" (PLC) can be applied to restore the missing content of packet-loss audio, so as to improve audio quality in real-time packetized audio communication applications.

With the renaissance of deep neural networks (DNNs), several approaches based on deep generative models have thrived in the PLC area, namely Deep PLC. By conducting supervised training on numerous clean-lossy audio pairs, a multitude of Deep PLC algorithms have demonstrated their superior capability in handling packet losses and predicting the future frames in a post-processing manner. In~{\cite{lee2016packet}}, by leveraging the spectral features of previous audio frames, a feed-forward neural network was proposed to predict the spectral features and then reconstruct the time-domain signal for each missing frame. Following the naive neural networks, a novel PLC framework based on recurrent neural network (RNN) showed the effectiveness of capturing more audio-related information in estimating lossy audio~{\cite{lotfidereshgi2018speech}}. Furthermore, by combining the usage of the long short-term memory networks (LSTM) and convolutional networks (CNNs), the convolutional recurrent network (CRN) for PLC~{\cite{lin2021time}} achieved remarkable performance in improving speech quality using the time-domain waveform mapping. More recently, on the account of its superiority in the audio generation, a generative adversarial network (GAN) was introduced to restore the missing part of the speech~{\cite{shi2019speech}} in PLC tasks. For example, a GAN-based approach~{\cite{pascual2021adversarial}} utilized mel-spectrogram of the lossy audio as input, and reconstructed the complete audio in the same way with the vocoders. Our preliminary work~{\cite{wang2021temporal}} considered both time-domain waveform and frequency-domain spectrum optimization, in other words, time-domain and frequency-domain discriminators were integrated into the classical GAN framework, leading to the improved PLC performance.

Motivated by the aforementioned Deep PLC studies, we aim to establish a low-complexity PLC framework, which can compensate for short-term losses totally and smooth over longer loss bursts. In this paper, following the typical GAN framework, we propose a novel generator inserted with gated vector quantizers, which learns to map the audio information to a high-level abstract space. On the basis of audio representation learning, we attach a gating mechanism to the vector quantizers to perform temporal collation of the encoding features representing the important content of speech. These modified vector quantizers enable the proposed network to "remember" the correlation of speech contexts like human memory, so as to reduce the impact of lost perturbations. Besides, an interleaved structure with the nested-UNet and the temporal feature-wise linear modulation (TFiLM) is utilized to assign temporal features for joint exploitation of long-term and short-term information correlation. Inspired by the preliminary study~{\cite{wang2021temporal}}, we integrate multi-resolution temporal discriminators and a complex-valued-spectrum discriminator to collaboratively provide external guidance in terms of the multi-scale waveform, spectral magnitude and phase, thus alleviating the time-frequency distortion. Comprehensive experimental results on the public dataset, namely INTERSPEECH 2022 Deep PLC challenge~{\cite{diener2022interspeech}}, show that the proposed method achieves competitive performance and outperforms many state-of-the-art PLC baselines.

\begin{figure*}[t]
	\centering
	\centerline{\includegraphics[width=1.95\columnwidth]{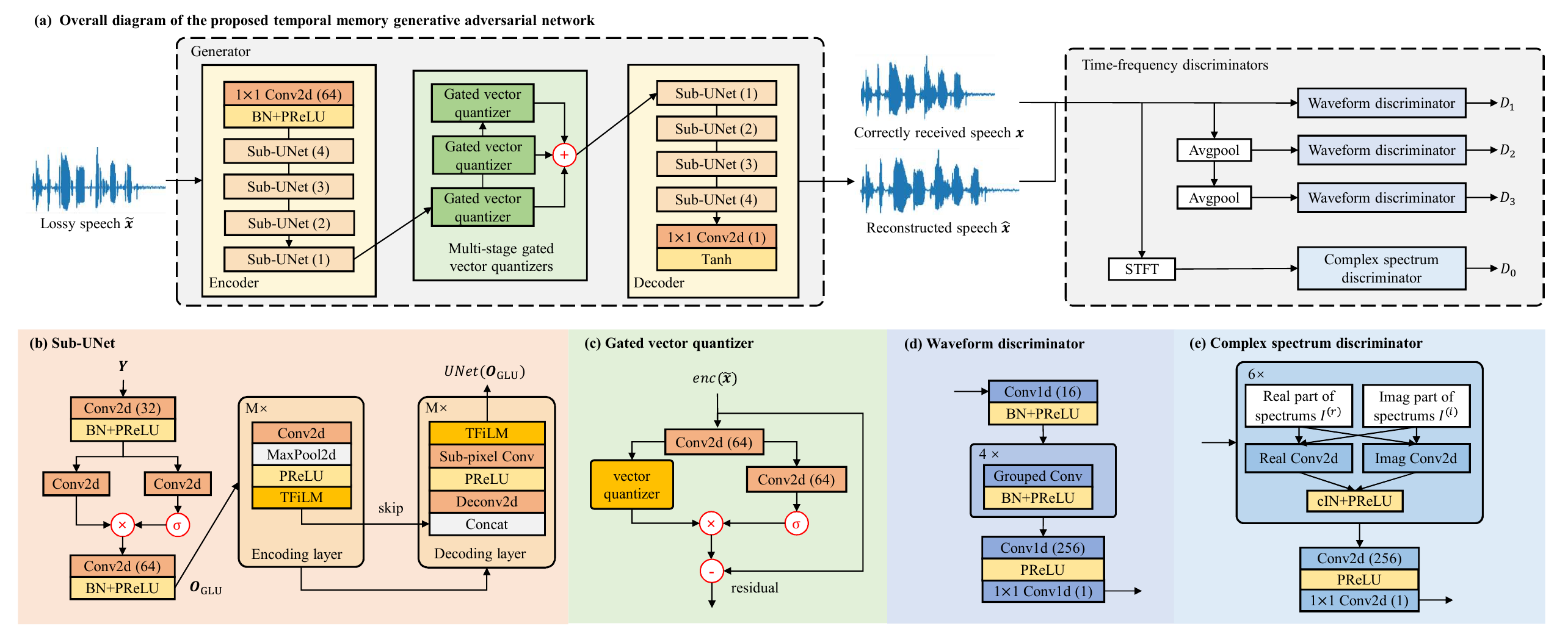}}
	\vspace{-0.3cm}
	\caption{The overall diagram of the proposed system. Different modules are indicated with different colors for better visualization. "skip" refers to the skip connection and $\sigma(\lambda)$ is the sigmoid function. $\left ( \bullet \right )$ after each convolutional layer represents the number of output channels for this convolution, while $\left ( \bullet \right )$ after each sub-UNet indicates the number of (de)encoding layers $M$ within this block.}
	\label{fig:networks}
	\vspace{-0.5cm}
\end{figure*}


\vspace{-0.35cm}
\section{Methodlogy\label{Section2}}
\vspace{-0.2cm}
\label{Sec2}
Following with the traditional PLC approach~{\cite{gunduzhan2001linear}}, the audio sequence decoded from a single packet is assumed to contain only one frame without loss of generality. For audio transmission, the continuous audio ${\bf x} \in \mathbb{R}^{1 \times L}$ is often split into short-time frames with the frame length $N$, and thus the $j$th audio frame ${\bf x}_j \in \mathbb{R}^{1 \times N}$ can be defined as:
\begin{equation}
	\label{eq:frame}
	x_j(n) = x \left ( N \cdot \left ( j-1 \right )+n \right ), n=0,\ldots,N-1,
\end{equation}
where $n$ is the time index of the $j$th frame, $L$ is the length of the input mixture and the number of frames $F$ is given by $F = \lfloor L/N\rfloor$. When the $j$th audio frame is lost, this frame becomes silent, and all observations in this frame are set to zero, i.e., ${\bf x}_j={\bf 0}^{1 \times N}$. We further define a received frame ${\bf \widetilde x}_j$ which can represent the clean state and lossy state of the $j$th frame simultaneously. With this definition, we have ${\widetilde{bf x}}_j = I_j{bf x}_j$, where $I_j$ becomes 1 when the $j$th frame is correctly received, otherwise $I_j = 0$. In a nutshell, the task for deep PLC is to map the received signal $\widetilde x(n)$ to the original high-quality audio $x(n)$. For real-time applications, the deep PLC needs to map the $j$th audio frame with the $j$th received audio frame and its previous received/mapped audio frames for causality. 

As illustrated in Figure~{\ref{fig:networks}}(a), the proposed system is mainly comprised of two sub-networks, namely a nested-UNet generator and the collaborative time-domain and frequency-domain discriminators. The nested-UNet generator aims at exporting the reconstructed frames ${\bf \widehat x}_j$ that can deceive the time-frequency integrated discriminators, while these discriminators attempt to find the best decision boundary between the reconstructed frames and the correctly received ones. By using adversarial training, the generated output is forced to be indistinguishable from the natural audio, and meanwhile multiple objective metrics can be optimized with different discriminators, such as the spectral details and voiced$/$unvoiced energy distribution. The detailed architectures of each module are illustrated below.
\vspace{-0.1cm}
\subsection{Nested-UNet generator \label{Section21}}
\vspace{-0.1cm}
Although the previous generative modeling studies~{\cite{kuleshov2017audio}} have verified the effectiveness of UNet-style architectures in the audio reconstruction task, the construction of deeper UNet architectures usually sacrifices high-resolution feature maps, which may cause the temporal information loss~{\cite{qin2020u2}} and degrade the quality of reconstructed audio. To tackle this shortcoming, we follow the architecture design patterns from the nested-UNet structure~{\cite{qin2020u2}}, and utilize a combination of the gated linear unit (GLU) and a sub-UNet in the proposed generator to mitigate the information loss and enhance the temporal dependencies, as presented in Figure~{\ref{fig:networks}}(b). 

Specifically, we employ a novel sub-UNet instead of the normal convolutional layers in original UNet to accomplish the multi-scale audio compression and restoration, which is a classical symmetric encoder-decoder structure. Then, before feeding the input features into each sub-Net, we employ GLUs to reduce the impact of missing information, which have been demonstrated its temporal modeling performance~{\cite{li2020time}}. Input feature map is gradually downsampled by multiple encoding layers, and the mirrored decoding layers reconstruct the encoded features to the original size. Compared with non-nested UNet, the stacking of proposed sub-UNets can leverage several contextual collations, progressive sampling and more filters with various receptive fields. More frequent sampling operations and the spatial resolution adjustments allow the proposed model to simulate finer reconstruction of lost information, and effectively extenuate the loss of waveform details caused by direct large-scale upsampling. 

In~{\cite{marafioti2020gacela}}, it has been revealed that the reconstruction of lost frames requires not only nearby contextual information but also the long-term information to improve the overall consistency of the reconstructed audio. In this regard, we employ the temporal feature-wise linear modulation (TFiLM)~{\cite{birnbaum2019temporal}} as the normalization layer of sub-UNet. By combining the pooling operations and LSTM, TFiLM can assist in capturing inter-frame dependencies along with frame-level inputs $Y \in \mathbb{R}^{C \times F \times N}$, which can be formulated as:
\begin{gather}
	\label{eq:tfilm}
	O_{\rm pool} = MaxPooling(Y), \\
	O_{\rm TFiLM} = LSTM(O_{\rm pool}) \cdot Y,
\end{gather}
where the pooling output $O_{\rm pool} \in \mathbb{R}^{C \times F \times 1}$ is obtained by the pooling operation along the temporal dimension $N$, allowing LSTM to focus on the information correlation refinement of the frame dimension $F$. Note that on the account of LSTMs invoking the pooling features, the computational efficiency of TFiLM layers still meets the requirement of real-time systems. 

Each encoding and decoding layer in a sub-UNet is comprised of plain convolution, TFiLM normalization and PReLU activation~{\cite{he2015delving}}. The downsampling and upsampling operations are based on max-pooling and sub-pixel convolution~{\cite{shi2016real}}, respectively. The number of (de)encoding layers $M$ in the sub-UNets are set to $\left \{4,3,2,1\right \}$ for the encoder and mirrored $\left \{1,2,3,4\right \}$ for the decoder. 

\vspace{-0.1cm}
\subsection{Gated vector quantizers \label{Section22}}
\vspace{-0.1cm}
In real-world communication scenarios, the potential large-scale loss bursts(\emph{e.g.} up to 1000 milliseconds) cause the difficulty to output normal speech with a small amount of useful context. Motivated by recent studies on Vector Quantization (VQ)~{\cite{van2017neural}}, we insert the vector quantizers into the proposed generator to capture the correct speech content of processing frames. To be specific, the vector quantizer maps the encoded features $enc({\bf \widetilde x})$ to discrete latent variables, and the decoder learns to reconstruct the smooth speech ${\bf \widehat x}$ with the learnable discrete codebook ${\bf Q}^V=\left \{ q_1,q_2,\ldots,q_V \right \}$, which can be formulated as:
\begin{equation}
	\begin{aligned}
		\label{eq:quantization}
		Quantize(enc({\bf \widetilde x}_j))=q_j, \\
		q_j = \mathop{\arg\min}\limits_{q \in {\bf Q}^V}(\left \| enc({\bf \widetilde x}_j) - q \right \|_{2}^{2}),
	\end{aligned}
\end{equation}
where $V$ is the codebook size and the function $Quantize$ selects the vector $q_j$ to be closest to $enc({\bf \widetilde X}^l)$ based on Euclidean distances. The subscript $j$ still denotes the frame index. 

Furthermore, owing to the fact that the recording of burst losses in PLC tasks may be detrimental to the codebook, we incorporate the gating mechanism into the original VQ architecture. As shown in Figure~{\ref{fig:networks}}(c), an additional gated branch is attached to the input layer of the modified vector quantizer, which consists of a plain convolutional layer and a sigmoid function. The gating mechanism enhances the contextual relevance of the latent codebook and picks out prioritized speech content for quantitative learning. In addition, we cascade multiple vector quantizers in a residual fashion~{\cite{vasuki2006review}} to learn more fine-grained ambiguous speech content in lossy speech. The unquantized input vector is passed through the first modified vector quantizer and the error between the original and quantized vector is then encoded with a sequence of additional quantizers, resulting in a progressively refined quantization of the input vector.  

The modified vector quantizer is dubbed as gated vector quantizer (GVQ), in which the kernel size and stride are set to (1,~3) and (1,~1), respectively. The number of gated vector quantizers is set to 3 and each quantizer uses a discrete codebook of size 512. To simplify the training process, we optimize each vector quantizer with exponential moving average k-means, similar to~{\cite{razavi2019generating}}. 

\vspace{-0.3cm}
\subsection{Time-frequency discriminators \label{Section23}}
\vspace{-0.2cm}
Inspired by the classification learning in multiple domains, two discriminators are adopted to separately learn the classification of the waveform and complex spectrum, which encourage the generator to produce reconstructed signals that are indistinguishable from correctly received audio. The diagrams of the proposed discriminators are shown in Figure~{\ref{fig:networks}}(d) and (e).

For the waveform discriminator, we utilize the multi-resolution temporal discriminator~{\cite{kumar2019melgan}}, in which three structurally identical discriminators are applied to the input audio at different resolutions. As described in~{\cite{kumar2019melgan}}, each single-scalar discriminator is mainly composed of an initial convolutional layer and four grouped convolutional layers, followed by batch normalization (BN)~{\cite{ioffe2015batch}} and PReLU. For the complex spectrum discriminator, it follows the design of our previous study~{\cite{wang2021temporal}}, in which complex-valued 2D convolutions are used for the classification of complex-valued spectra and promote, resulting in fine-grained prediction of both the magnitude and phase of the lost frames. The complex instance normalization (IN) and complex PReLU~{\cite{hu2020dccrn}} operate on both real and imaginary values. Finally, a $1 \times 1$ convolutional layer is employed in both waveform and complex spectrum discriminators to aggregate the corresponding discrimination results.

\vspace{-0.3cm}
\subsection{Loss function \label{Section24}}
\vspace{-0.2cm}
During the adversarial training process, we adopt a combination of the adversarial loss and the feature loss. The adversarial loss is used to establish the competitive relationship between discriminators and the generator. Similar to~{\cite{baby2019sergan}}, we adopt the relativistic average least-square loss to stabilize the adversarial training, in which the relativistic logit aggregates the means of multiple discriminative results and reduces the discrepancy between reconstructed audio ${\bf \widehat x}$ and correctly received audio ${\bf x}$, which can be formulated as:
\begin{equation}
	\bar{D}_{{\bf x}} = \frac{1}{K}\sum_{k}D_k({\bf x})-\mathbb{E}_{{\bf \widehat x}} \left [\frac{1}{K}\sum_{k}D_k({\bf \widehat x})  \right ],
\end{equation}

\begin{equation}
	\bar{D}_{{\bf \widehat x}} = \frac{1}{K}\sum_{k}D_k({\bf \widehat x})-\mathbb{E}_{{\bf  x}} \left [\frac{1}{K}\sum_{k}D_k({\bf x}) \right ],
\end{equation}
where $D_k({\bf x})$ and $D_k({\bf \widehat x})$ represent the output vector of $k$th discriminator with the correctly received frames ${\bf x}$ and the reconstructed frames ${\bf \widehat x}$ as input, respectively. Note that $k \in \left \{ 0,1,\dots,K \right \}$ denotes the index over the individual discriminators, where $k=0$ presents the complex spectrum discriminator and $k \in \left \{ 1,\dots,K \right \}$ presents the different resolutions of the waveform discriminators ($K=3$ in this paper). Hence, the loss function of the discriminator and its corresponding generator can be expressed as below:
\begin{equation}
	\mathcal{L}_{D} = \mathbb{E}_{{\bf x}} [(\bar{D}_{{\bf x}}-1)^2] + \mathbb{E}_{{\bf \widehat x}} [(\bar{D}_{{\bf \widehat x}}+1)^2].
\end{equation}

\begin{equation}
	\mathcal{L}_{G}^{\rm adv} = \mathbb{E}_{{\bf \widehat x}} [(\bar{D}_{{\bf \widehat x}}-1)^2] + \mathbb{E}_{{\bf x}} [(\bar{D}_{{\bf x}}+1)^2],
\end{equation}

For the feature loss, we incorporate the Mean Squared Error (MSE) loss and the multi-scale spectral reconstruction loss~{\cite{yamamoto2020parallel}} to facilitate the audio fidelity and perceptual quality. Given the inpainted time-domain signal ${\bf \widehat x}$ and the target one ${\bf x}$ as well as their corresponding short-time Fourier transforms (STFT) $\widehat S$ and $S$, the feature loss can be given by: 
\begin{gather}
	\mathcal{L}_{G}^{\rm mse}= \left \| {\bf x}-  {\bf \widehat x}\right \|^{2}_{F},\\
	\mathcal{L}_{G}^{\rm spec}= \frac{\left \| \left | S(l,k) \right |-\left | \widehat S(l,k) \right | \right \|_{F}}{\left \| \left | S(l,k) \right | \right \|_{F}}+\frac{1}{L}\left \| log \frac{\left |S(l,k) \right |}{\left |\widehat S(l,k) \right |} \right \|_{1},
\end{gather}
where $l$ and $k$ denote the frame index and the frequency index, respectively. The multi-resolution STFT loss is the summation of the STFT losses with different lengths of window to improve the frequency robustness of the output speech. 

Therefore, the full loss for the generator is given by the weighted sum of all the above mentioned loss components:
\begin{equation}
	\mathcal{L}_{G} = \mathcal{L}_{G}^{\rm adv}+\lambda_{\rm mse} \mathcal{L}_{G}^{\rm mse}+\lambda_{\rm spec} \mathcal{L}_{G}^{\rm spec},
\end{equation}
where $\lambda_{\rm mse}$ and $\lambda_{\rm spec}$ are the weighting hyper-parameters, which are initially set to be 1.0 and 0.5, respectively, aiming to balance the learning rate of each loss. 

\renewcommand\arraystretch{1.1}
\begin{table*}[t]
	\centering
	\footnotesize
	\setlength\tabcolsep{1.5pt}
	\caption{Comparison with other state-of-art systems on the PLC Challenge non-blind test set. \textbf{BOLD} indicates the highest performance}
	\vspace{-0.35cm}
	\scalebox{0.9}{
		\begin{tabular}{cc|cccc|ccc|ccc|ccc}
			\hline\hline
			&\textbf{Models}  &Domain &Param.(M) &MACs(G/s) &\multicolumn{1}{c|}{RTF}  &\multicolumn{3}{c|}{\textbf{WB-PESQ}}  &\multicolumn{3}{c|}{\textbf{STOI($\%$)}} &\multicolumn{3}{c}{\textbf{PLCMOS}} \\
			\cline{1-15}
			& \textbf{Burst subset} &- &- &- &- &(0, 120] &(120,320] &\multicolumn{1}{c|}{(320,1000]} &(0, 120] &(120,320] &\multicolumn{1}{c|}{(320,1000]} &(0, 120] &(120,320] &\multicolumn{1}{c}{(320,1000]} \\
			\cline{1-15}
			&\multicolumn{1}{c|}{\textbf{Lossy speech}} &- &- &- &- &2.59 &1.75 &\multicolumn{1}{c|}{1.72} &89.42 &70.83 &\multicolumn{1}{c|}{65.02} &3.13 &2.47 &\multicolumn{1}{c}{2.82} \\
			&\multicolumn{1}{c|}{\textbf{NSNet2}} &- &- &- &- &2.39 &1.71 &\multicolumn{1}{c|}{1.67} &84.82 &64.83 &\multicolumn{1}{c|}{60.96} &3.08 &2.39 &\multicolumn{1}{c}{2.61} \\
			&\multicolumn{1}{c|}{\textbf{CRN}} &T &4.92 &0.52 &0.57 &2.63 &1.78 &\multicolumn{1}{c|}{1.74} &89.64 &72.70 &\multicolumn{1}{c|}{67.27} &3.26 &2.55 &\multicolumn{1}{c}{2.49} \\
			&\multicolumn{1}{c|}{\textbf{TFNet}} &T-F &2.55 &1.59 &0.88  &2.90 &1.93 &\multicolumn{1}{c|}{1.81} &91.92 &78.02 &\multicolumn{1}{c|}{69.93} &3.88 &2.80 &\multicolumn{1}{c}{2.59} \\
			&\multicolumn{1}{c|}{\textbf{TFGAN}} &T-F &1.85 &0.71 &0.57 &3.03 &2.14 &\multicolumn{1}{c|}{1.86} &92.32 &78.27 &\multicolumn{1}{c|}{70.16} &3.99 &3.07 &\multicolumn{1}{c}{2.66} \\
			&\multicolumn{1}{c|}{\textbf{TMGAN}} &T-F &2.85 &2.12 &0.74 &\textbf{3.37} &\textbf{2.33} &\multicolumn{1}{c|}{\textbf{2.08}} &\textbf{93.37} &\textbf{79.50} &\multicolumn{1}{c|}{\textbf{71.56}} &\textbf{4.57} &\textbf{4.28} &\multicolumn{1}{c}{\textbf{4.21}} \\
			\hline\hline
	\end{tabular}}		
	\label{tab:comparison}
	\vspace{-0.5cm}
\end{table*}

\vspace{-0.1cm}
\section{Experiments\label{Section3}}
\label{Sec3}
\vspace{-0.1cm}
\subsection{Datasets\label{Section31}}
\vspace{-0.1cm}
To evaluate the performance of our framework, we conduct extensive experiments on the INTERSPEECH 2022 Audio Deep Packet Loss Concealment Challenge dataset~{\cite{diener2022interspeech}}, which consists of 23,184 clean-lossy pairs for training. The Microsoft team created this dataset, where each clip of audio was sampled from a base public domain podcast dataset with the duration about 10 seconds. All the lossy audio clips were obtained by zeroing out the corresponding regions according to the losses from the sampled traces~{\cite{diener2022interspeech}}. Specifically, the lossy utterances can be divided into three subsets according to the maximum burst loss length in the trace: $\left\{(0, 120] ms, (120, 320] ms, (320, 1000] ms\right\}$. For the test set, we utilize the test dataset of PLC-Challenge, which consists of 966 real recordings in the English language. 

\vspace{-0.2cm}
\subsection{Implementation setup\label{Section32}}
\vspace{-0.1cm}
To leverage the gated vector quantizer to learn more speech content, we incorporate additional clean utterances from the VCTK corpus~{\cite{veaux2013vbk-dataset}} in the training process. All clips of audio are sampled at 16 kHz and the length of each frame is set to 20 ms. The proposed model and baselines are trained for 100 epochs using the RAdam optimizer~{\cite{liu2020radam}} with a learning rate of 0.0001, and the corresponding attenuation rates are $\beta_{1}=0.9$ and $\beta_{2}=0.999$. To stabilize the training process, the learning rates of the discriminators are set to 0.00005. The generator directly extracts the features from the original waveform, while the complex spectrum discriminator needs to perform 320-point FFT on waveform beforehand. The training batch size is set to 8 at the utterance level. To further smooth the transition between reconstructed frames and correctly received frames, we introduce a speech enhancement post-processing module according to our preliminary studies.~{\cite{li2021simultaneous}}.

\vspace{-0.1cm}
\section{Experimental results and discussion\label{Section4}}
\label{Sec4}		
\vspace{-0.1cm}
In this study, we use the wide-band perceptual evaluation of speech quality (WB-PESQ)~{\cite{rix2001perceptual}}, short-time objective intelligibility (STOI)~{\cite{taal2010short}}, word error rate (WER) and PLCMOS~{\cite{diener2022interspeech}} to evaluate PLC performance of the proposed model. Higher values indicate better performance. 

\vspace{-0.2cm}  
\subsection{Ablation study \label{Section41}}
\vspace{-0.1cm}  
For ablation studies on the proposed model, we reassemble the above modules and calculate the average evaluation scores of each scheme on the test set. We first train two UNet-style architectures with the guidance of mixed time/frequency-domain discriminators, named traditional UNet and nested-UNet. As shown in Table~{\ref{tab:ablation-study}}, the nested structure achieves significant improvements than traditional UNet in terms of all metrics, especially in improving STOI and PLCMOS scores. Subsequently, the combination of the nested-UNet and TFiLM achieves consistently better performance, indicating that the memory mechanism of LSTM can effectively improve the reconstruction quality of lossy speech. Finally, when incorporating the single raw vector quantizer$/$multi-stage gated vector quantizers into the nested-UNet, one can observe that the naive VQ affected by the burst losses even degrades speech quality, while GVQ demonstrates its superiority in capturing correct speech content.

\vspace{-0.1cm}
\subsection{Comparison with SOTA methods\label{Section42}}
\vspace{-0.1cm}
In different burst-loss length subsets, we further compare our model with several state-of-the-art PLC methods, including the PLC-Challenge baseline NSNet2~{\cite{diener2022interspeech}}, CRN~{\cite{lin2021time}}, TFNet~{\cite{jiang2022end}} and our earlier proposed model TFGAN~{\cite{wang2021temporal}}. From Table~{\ref{tab:comparison}}, several observations can be obtained. First, by introducing the spectral features, all the mixed time/frequency methods outperform CRN by a large margin, which only relies on the time-domain information. Second, when comparing TFNet with TFGAN, one can observe that transferring the time-frequency mapping task to discriminators does not degrade the reconstructed speech quality. This operation simplifies the inference procedure of the generator, leading to less real-time processing time (RTF). Third, compared with other advanced PLC methods, the proposed method achieves relatively better performance in terms of all metrics. Furthermore, in the large-scale burst subset of $(320, 1000] ms$, the proposed method significantly surpasses other baselines in terms of the PLCMOS score, demonstrating the remarkable reconstruction performance of TMGAN. In Table~{\ref{tab:ablation-study}}, we present the subjective results of the proposed method in terms of PLCMOS~{\cite{diener2022interspeech}}, DNSMOS~{\cite{reddy2021dnsmos}} and Crowd-Sourced Mean Opinion Score (CMOS)~{\cite{diener2022interspeech}} on the PLC Challenge blind test set. Compared with the zero-filling baseline, TMGAN provides significant improvements in terms of overall speech quality. The number of trainable parameters of the proposed framework is 2.85 million, and the number of multiply-accumulate operations (MACs) per second is 2.12 G as shown in Table~{\ref{tab:plc-challenge}}. The one-frame processing time of our system with the PyTorch implementation is around 14.8 ms on an Intel i5-4300U PC. With two frames looking ahead (40ms) and the strade time (20 ms), the algorithm latency of proposed model is about 80 ms.

\renewcommand\arraystretch{1.1}
\begin{table}[t]
	\centering
	\footnotesize
	\setlength\tabcolsep{3.0pt}
	\caption{Ablation study on the proposed method. “$+$TFiLM” denotes the introduction of temporal feature-wise linear modulations. “$+$VQ” denotes that a raw vector quantizer is incorporated, while “$+$GVQ” represents the multi-stage gated vector quantizers.}
	\vspace{-0.3cm}
	\scalebox{0.9}{
		\begin{tabular}{cc|ccc}
			\hline\hline
			&\textbf{Models} &\textbf{WB-PESQ}  &\textbf{STOI($\%$)} &\textbf{PLCMOS} \\
			\cline{1-5}
			&\multicolumn{1}{c|}{\textbf{Traditional UNet}} &2.44 &82.24 &3.25 \\
			&\multicolumn{1}{c|}{\textbf{Nested-UNet}} &2.59 &84.37 &3.54 \\
			&\multicolumn{1}{c|}{\textbf{~ + TFiLM}} &2.73 &84.91 &4.04 \\
			&\multicolumn{1}{c|}{\textbf{~~ + VQ}} &2.65 &84.02 &3.86 \\
			&\multicolumn{1}{c|}{\textbf{~~ + GVQ}} &\textbf{2.83} &\textbf{85.47} &\textbf{4.14} \\
			\hline\hline
	\end{tabular}}		
	\label{tab:ablation-study}
	\vspace{-0.3cm}
\end{table}

\renewcommand\arraystretch{1.1}
\begin{table}[t]
	\centering
	\footnotesize
	\setlength\tabcolsep{3.0pt}
	\caption{Subjective evaluation on PLC Challenge blind test set.}
	\vspace{-0.3cm}
	\scalebox{0.9}{
		\begin{tabular}{cc|ccc}
			\hline\hline
			&\textbf{models} &\textbf{PLCMOS}  &\textbf{DNSMOS} &\textbf{CMOS} \\
			\cline{1-5}
			&\multicolumn{1}{c|}{\textbf{Zero-filling baseline}} &2.904 &3.444 &-1.231 \\
			&\multicolumn{1}{c|}{\textbf{TMGAN}} &\textbf{4.406} &\textbf{3.958} &\textbf{-0.279} \\
			\hline\hline
	\end{tabular}}		
	\label{tab:plc-challenge}
	\vspace{-0.6cm}
\end{table}

\vspace{-0.1cm}
\section{Conclusions\label{Section5}}
\label{Sec5}
\vspace{-0.1cm}

This paper proposes a novel packet loss concealment model based on temporal memory generative adversarial network (TMGAN-PLC). Taking a nested-UNet and the temporal feature-wise linear modulation as the backbone, the proposed generator can aggregate local and global features by leveraging the memory function in LSTMs, thus reducing the effect of buffer size on the reconstruction quality. Furthermore, we design the multi-stage vector quantizers with a gating mechanism to deal with burst losses. In the gated vector quantizers, the encoding learning can autonomously avoid missing the large regions, and the decoding reconstruction incorporates the codebook and quantization features of previous frames, which can recover the high-quality audio even for long burst losses. Experimental results demonstrate the effectiveness of each temporal memory module, and show the competitive performance of the proposed method in terms of speech quality, intelligibility and PLCMOS. 

\vfill\pagebreak
\bibliographystyle{IEEEtran}
\bibliography{myrefs}

\begin{thebibliography}{10}
\providecommand{\url}[1]{#1}
\csname url@samestyle\endcsname
\providecommand{\newblock}{\relax}
\providecommand{\bibinfo}[2]{#2}
\providecommand{\BIBentrySTDinterwordspacing}{\spaceskip=0pt\relax}
\providecommand{\BIBentryALTinterwordstretchfactor}{4}
\providecommand{\BIBentryALTinterwordspacing}{\spaceskip=\fontdimen2\font plus
\BIBentryALTinterwordstretchfactor\fontdimen3\font minus
  \fontdimen4\font\relax}
\providecommand{\BIBforeignlanguage}[2]{{%
\expandafter\ifx\csname l@#1\endcsname\relax
\typeout{** WARNING: IEEEtran.bst: No hyphenation pattern has been}%
\typeout{** loaded for the language `#1'. Using the pattern for}%
\typeout{** the default language instead.}%
\else
\language=\csname l@#1\endcsname
\fi
#2}}
\providecommand{\BIBdecl}{\relax}
\BIBdecl

\bibitem{takahashi2004packet}
A.~Takahashi, H.~Yoshino, and N.~Kitawaki, ``Perceptual {QoS} assessment
  technologies for {VoIP},'' \emph{IEEE Communications Magazine}, vol.~42,
  no.~7, pp. 28--34, 2004.

\bibitem{lee2016packet}
B.~K. Lee and J.~H. Chang, ``Packet loss concealment based on deep neural
  networks for digital speech transmission,'' \emph{IEEE/ACM Transactions on
  Audio Speech and Language Processing (TASLP)}, vol.~24, no.~2, pp. 378--387,
  2016.

\bibitem{lotfidereshgi2018speech}
R.~Lotfidereshgi and P.~Gournay, ``Speech prediction using an adaptive
  recurrent neural network with application to packet loss concealment,'' in
  \emph{2018 IEEE International Conference on Acoustics, Speech and Signal
  Processing (ICASSP)}.\hskip 1em plus 0.5em minus 0.4em\relax IEEE, 2018, pp.
  5394--5398.

\bibitem{lin2021time}
J.~Lin, Y.~Wang, K.~Kalgaonkar, G.~Keren, D.~Zhang, and C.~Fuegen, ``A
  time-domain convolutional recurrent network for packet loss concealment,'' in
  \emph{2021 IEEE International Conference on Acoustics, Speech and Signal
  Processing (ICASSP)}.\hskip 1em plus 0.5em minus 0.4em\relax IEEE, 2021, pp.
  7148--7152.

\bibitem{shi2019speech}
Y.~Shi, N.~Zheng, Y.~Kang, and W.~Rong, ``Speech loss compensation by
  generative adversarial networks,'' in \emph{2019 Asia-Pacific Signal and
  Information Processing Association Annual Summit and Conference (APSIPA
  ASC)}.\hskip 1em plus 0.5em minus 0.4em\relax IEEE, 2019, pp. 347--351.

\bibitem{pascual2021adversarial}
S.~Pascual, J.~Serr{\`a}, and J.~Pons, ``Adversarial auto-encoding for packet
  loss concealment,'' in \emph{2021 IEEE Workshop on Applications of Signal
  Processing to Audio and Acoustics (WASPAA)}.\hskip 1em plus 0.5em minus
  0.4em\relax IEEE, 2021, pp. 71--75.

\bibitem{wang2021temporal}
J.~Wang, Y.~Guan, C.~Zheng, R.~Peng, and X.~Li, ``A temporal-spectral
  generative adversarial network based end-to-end packet loss concealment for
  wideband speech transmission,'' \emph{The Journal of the Acoustical Society
  of America}, vol. 150, no.~4, pp. 2577--2588, 2021.

\bibitem{diener2022interspeech}
L.~Diener, S.~Sootla, S.~Branets, A.~Saabas, R.~Aichner, and R.~Cutler,
  ``Interspeech 2022 audio deep packet loss concealment challenge,'' in
  \emph{{INTERSPEECH} 2022 - 23rd Annual Conference of the International Speech
  Communication Association}, 2022 (submitted).

\bibitem{gunduzhan2001linear}
E.~Gunduzhan and K.~Momtahan, ``Linear prediction based packet loss concealment
  algorithm for {PCM} coded speech,'' \emph{IEEE Transactions on Speech and
  Audio Processing}, vol.~9, no.~8, pp. 778--785, 2001.

\bibitem{kuleshov2017audio}
V.~Kuleshov, S.~Z. Enam, and S.~Ermon, ``Audio super-resolution using neural
  nets,'' in \emph{the Fifth International Conference on Learning
  Representations (ICLR 2017-Workshop Track)}, 2017.

\bibitem{qin2020u2}
X.~Qin, Z.~Zhang, C.~Huang, M.~Dehghan, O.~R. Zaiane, and M.~Jagersand,
  ``U2-net: Going deeper with nested u-structure for salient object
  detection,'' \emph{Pattern Recognition}, vol. 106, p. 107404, 2020.

\bibitem{li2020time}
A.~Li, C.~Zheng, L.~Cheng, R.~Peng, and X.~Li, ``A time-domain monaural speech
  enhancement with feedback learning,'' in \emph{2020 Asia-Pacific Signal and
  Information Processing Association Annual Summit and Conference (APSIPA
  ASC)}.\hskip 1em plus 0.5em minus 0.4em\relax IEEE, 2020, pp. 769--774.

\bibitem{marafioti2020gacela}
A.~Marafioti, P.~Majdak, N.~Holighaus, and N.~Perraudin, ``Gacela: A generative
  adversarial context encoder for long audio inpainting of music,'' \emph{IEEE
  Journal of Selected Topics in Signal Processing}, vol.~15, no.~1, pp.
  120--131, 2020.

\bibitem{birnbaum2019temporal}
S.~Birnbaum, V.~Kuleshov, Z.~Enam, P.~W.~W. Koh, and S.~Ermon, ``Temporal film:
  Capturing long-range sequence dependencies with feature-wise modulations.''
  \emph{Advances in Neural Information Processing Systems (NeurIPS 2019)},
  vol.~32, pp. 10\,287--10\,298, 2019.

\bibitem{he2015delving}
K.~He, X.~Zhang, S.~Ren, and J.~Sun, ``Delving deep into rectifiers: Surpassing
  human-level performance on imagenet classification,'' in \emph{Proceedings of
  the 2015 IEEE International Conference on Computer Vision (ICCV)}, 2015, pp.
  1026--1034.

\bibitem{shi2016real}
W.~Shi, J.~Caballero, F.~Husz{\'a}r, J.~Totz, A.~P. Aitken, R.~Bishop,
  D.~Rueckert, and Z.~Wang, ``Real-time single image and video super-resolution
  using an efficient sub-pixel convolutional neural network,'' in
  \emph{Proceedings of the IEEE conference on computer vision and pattern
  recognition (CVPR)}, 2016, pp. 1874--1883.

\bibitem{van2017neural}
A.~van~den Oord, O.~Vinyals, and K.~Kavukcuoglu, ``Neural discrete
  representation learning,'' in \emph{Proceedings of the 31st International
  Conference on Neural Information Processing Systems (NIPS'17)}, California,
  USA, 2017, pp. 6309--6318.

\bibitem{vasuki2006review}
A.~Vasuki and P.~Vanathi, ``A review of vector quantization techniques,''
  \emph{IEEE Potentials}, vol.~25, no.~4, pp. 39--47, 2006.

\bibitem{razavi2019generating}
A.~Razavi, A.~van~den Oord, and O.~Vinyals, ``Generating diverse high-fidelity
  images with vq-vae-2,'' vol.~32, Vancouver, Canada, 2019, pp.
  14\,866--14\,876.

\bibitem{kumar2019melgan}
K.~Kumar, R.~Kumar, T.~de~Boissiere, L.~Gestin, W.~Z. Teoh, J.~Sotelo,
  A.~de~Br{\'e}bisson, Y.~Bengio, and A.~C. Courville, ``Melgan: Generative
  adversarial networks for conditional waveform synthesis,'' \emph{Advances in
  Neural Information Processing Systems (NeurIPS 2019)}, vol.~32, 2019.

\bibitem{ioffe2015batch}
S.~Ioffe and C.~Szegedy, ``Batch normalization: Accelerating deep network
  training by reducing internal covariate shift,'' in \emph{Proceedings of the
  32nd International Conference on Machine Learning}.\hskip 1em plus 0.5em
  minus 0.4em\relax PMLR, 2015, pp. 448--456.

\bibitem{hu2020dccrn}
Y.~Hu, Y.~Liu, S.~Lv, M.~Xing, S.~Zhang, Y.~Fu, J.~Wu, B.~Zhang, and L.~Xie,
  ``{DCCRN}: Deep complex convolution recurrent network for phase-aware speech
  enhancement,'' \emph{Proceedings of the 21th Annual Conference of the
  International Speech Communication Association (Interspeech 2020)}, pp.
  2472--2476, 2020.

\bibitem{baby2019sergan}
D.~Baby and S.~Verhulst, ``{SERGAN}: Speech enhancement using relativistic
  generative adversarial networks with gradient penalty,'' in \emph{2019 IEEE
  International Conference on Acoustics, Speech and Signal Processing
  (ICASSP)}.\hskip 1em plus 0.5em minus 0.4em\relax IEEE, 2019, pp. 106--110.

\bibitem{yamamoto2020parallel}
R.~Yamamoto, E.~Song, and J.-M. Kim, ``Parallel {WaveGAN}: A fast waveform
  generation model based on generative adversarial networks with
  multi-resolution spectrogram,'' in \emph{2020 IEEE International Conference
  on Acoustics, Speech and Signal Processing (ICASSP)}.\hskip 1em plus 0.5em
  minus 0.4em\relax IEEE, 2020, pp. 6199--6203.

\bibitem{veaux2013vbk-dataset}
C.~Veaux, J.~Yamagishi, and S.~King, ``The voice bank corpus: Design,
  collection and data analysis of a large regional accent speech database,'' in
  \emph{2013 international conference oriental COCOSDA held jointly with 2013
  conference on Asian spoken language research and evaluation
  (O-COCOSDA/CASLRE)}.\hskip 1em plus 0.5em minus 0.4em\relax IEEE, 2013, pp.
  1--4.

\bibitem{liu2020radam}
L.~Liu, H.~Jiang, P.~He, W.~Chen, X.~Liu, J.~Gao, and J.~Han, ``On the variance
  of the adaptive learning rate and beyond,'' in \emph{the Eighth International
  Conference on Learning Representations (ICLR 2020)}, Addis Ababa, Ethiopia,
  2020.

\bibitem{li2021simultaneous}
A.~Li, W.~Liu, X.~Luo, G.~Yu, C.~Zheng, and X.~Li, ``A simultaneous denoising
  and dereverberation framework with target decoupling,'' \emph{Proceedings of
  the 22th Annual Conference of the International Speech Communication
  Association (Interspeech 2021)}, pp. 2801--2805, 2021.

\bibitem{rix2001perceptual}
A.~W. Rix, J.~G. Beerends, M.~P. Hollier, and A.~P. Hekstra, ``Perceptual
  evaluation of speech quality ({PESQ})-a new method for speech quality
  assessment of telephone networks and codecs,'' in \emph{2001 IEEE
  International Conference on Acoustics, Speech, and Signal Processing
  (ICASSP)}, vol.~2.\hskip 1em plus 0.5em minus 0.4em\relax IEEE, 2001, pp.
  749--752.

\bibitem{taal2010short}
C.~H. Taal, R.~C. Hendriks, R.~Heusdens, and J.~Jensen, ``A short-time
  objective intelligibility measure for time-frequency weighted noisy speech,''
  in \emph{2010 IEEE international conference on acoustics, speech and signal
  processing (ICASSP)}.\hskip 1em plus 0.5em minus 0.4em\relax IEEE, 2010, pp.
  4214--4217.

\bibitem{jiang2022end}
X.~Jiang, X.~Peng, C.~Zheng, H.~Xue, Y.~Zhang, and Y.~Lu, ``End-to-end neural
  audio coding for real-time communications,'' \emph{arXiv preprint
  arXiv:2201.09429}, 2022.

\bibitem{reddy2021dnsmos}
C.~K. Reddy, V.~Gopal, and R.~Cutler, ``Dnsmos: A non-intrusive perceptual
  objective speech quality metric to evaluate noise suppressors,'' in
  \emph{ICASSP 2021-2021 IEEE International Conference on Acoustics, Speech and
  Signal Processing (ICASSP)}.\hskip 1em plus 0.5em minus 0.4em\relax IEEE,
  2021, pp. 6493--6497.

\end{thebibliography}
	
\end{document}